\documentclass[nofootinbib,onecolumn]{revtex4}
\usepackage{graphicx,color}
\usepackage{amssymb}
\usepackage{amsmath}
\usepackage{bm}
\usepackage{latexsym}
\usepackage{dcolumn}
\usepackage{float}
\usepackage{hyperref}

\begin{document}

\title{Cosmology from a new non-conservative gravity}

\author{J\'ulio C. Fabris}\email{julio.fabris@cosmo-ufes.org}
\affiliation{Universidade Federal do Esp\'{\i}rito Santo (UFES), Av. Fernando Ferrari S/N, 29075-910, Vit\'oria, Brazil}

\affiliation{National Research Nuclear University “MEPhI”,
Kashirskoe sh.  31, Moscow 115409, Russia}
\author{Hermano Velten}\email{velten@pq.cnpq.br}
\author{Thiago R. P. Caram\^es}\email{trpcarames@gmail.com}
\affiliation{Universidade Federal do Esp\'{\i}rito Santo (UFES), Av. Fernando Ferrari S/N, 29075-910, Vit\'oria, Brazil}
\author{Matheus J. Lazo}\email{matheusjlazo@gmail.com}
\affiliation{Instituto  de Matem\'atica,  Estat\'istica  e F\'isica – FURG, Rio Grande,  RS, Brazil.}
\author{Gast\~ao S. F. Frederico}\email{gastao.frederico@ua.pt}
\affiliation{Departamento de Matem\'atica, Universidade Federal de Santa Catarina, Florian\'opolis, SC, Brazil}
\affiliation{Department of Science and Technology,  University of Cape Verde,  Praia, Cabo Verde}

\begin{abstract}
In this paper we present a cosmological model arising from a non-conservative gravitational theory proposed in \cite{lazo}. The novel feature where comparing with previous implementations of dissipative effects in gravity is the possible arising of such phenomena from a least action principle, so they are of a purely geometric nature. We derive the dynamical equations describing the behaviour of the cosmic background, considering a single fluid model composed by pressureles matter, whereas the dark energy is conceived as an outcome of the ``geometric'' dissipative process emerging in the model. Besides, adopting the synchronous gauge we obtain the first-order perturbative equations which shall describe the evolution of the matter perturbations within the linear regime.
\end{abstract}

\maketitle

\section{Introduction}

The first decades of the XXth century witnessed the ascension of the general relativity (GR) as the revolutionary paradigm for the gravitational interaction. Among all possible adjectives that can be assigned to GR, it is clearly a simple theory. Simple, in the sense that its field equations are obtained from the standard variational principle where the gravitational Lagrangian equals the Ricci scalar $\cal{L}\propto R$ which is the simplest Lagrangian in four dimensions (up to a cosmological constant term) leading to second order differential equations.

Although GR remains being considered the standard gravitational theory the observations of galaxy rotation curves and the inference of the accelerated expansion of the universe lead to the concept of dark matter and dark energy phenomena, respectively. The conclusion from these facts is that either the energy content of the universe is composed by strange forms of particles/fields or GR fails in describing the dynamics of galaxies and other cosmological observables. The latter assumption has led to the construction of several alternatives to GR. Most of them based on the fact that the gravitational Lagrangian has a non trivial dependence on geometrical quantities or even matter fields. There are also theories which relax some of the fundamental pillars over which GR has been built.

In this work we focus on the cosmological aspects of a recent gravitational theory proposed in \cite{lazo} in which dissipative processes are incorporated in the gravity by means of a generalization of the least action principle. This new theory of gravity is discussed in the next section. In section III we explore the cosmological scenario emerging from the modified field equations and derive the dynamical equations at background and perturbative levels. The results are compared either with the usual viscous fluid formulation and with the $\Lambda$CDM model.
In the section IV we bring a discussion on the perturbative aspect of such cosmology using the synchronous gauge. Lastly, the section V is dedicated to our concluding remarks.

\section{General equations}

A geometrical viscous gravity model can be obtained from first principles by an action-dependent Lagrangian formulation \cite{lazo}. A generalization of the action Principle for Action-dependent Lagrangians was introduced for the first time in the 30's by Herglotz in order to give a variational principle to dissipative phenomena \cite{her1,her2}. More precisely, the original Herglotz variational problem consists in the problem of determining the path $x(t)$ that extremize an Action of the form
\begin{eqnarray}
\label{H}
S = \int {\cal L}(x, \dot x, S) d t.
\end{eqnarray}
Herglotz proved \cite{her1,her2} that a necessary condition for a path $x(t)$ to be an extremizer of the variational problem (\ref{H}) is given by the generalized Euler-Lagrange equation
\begin{eqnarray}
\label{HEL}
\frac{d}{dt}\frac{\partial {\cal L}}{\partial \dot x} - \frac{\partial {\cal L}}{\partial x}  - \frac{\partial{\cal L}}{\partial S}\frac{\partial {\cal L}}{\partial \dot x} = 0.
\end{eqnarray}
The application of Herglotz problem to non-conservative systems is evident even in the simplest case where the dependence of the Lagrangian function on the Action is linear. For example, the Lagrangian ${\cal L}=\frac{m\dot{x}^2}{2}-U(x)-\frac{\gamma}{m} S$ describes a particle under viscous forces and, from (\ref{HEL}), the resulting equation of motion includes the well known dissipative force $\gamma \dot{x}$. In this context, the term linear on $S$ in the Lagrangian can be interpreted as a potential function for the non-conservative force. 

However, despite the Herglotz problem was introduced in 1930, a covariant generalization of (\ref{H}) for several variables was obtained only recently \cite{lazo}. The Lagrangian including the gravity sector considered in \cite{lazo} is given by
\begin{eqnarray}
{\cal L} = \sqrt{-g}(R - \lambda_{\mu}s^\mu) + {\cal L}_m,
\end{eqnarray}
where $s^\mu$ is an action-density field and $\lambda_\mu$ is coupling term which may depend on the space-time coordinate. The development of this proposal is given in details in Ref. \cite{lazo}. The inclusion of term $\lambda_{\mu}s^{\mu}$ can be seen as a covariant implementation of the linear dependence on the action approached by Herglotz in the classical mechanics context. So that, it is expected that such a term carries the dissipative nature of the theory. In this sense, the introduction of the two four-vectors $\lambda^{\mu}$ and $s^{\mu}$ revealed the most simple way to perform such generalization, and even a natural choice as a careful look at \cite{lazo} can show us. Notice that in that reference the proposal of the authors consisted in providing a covariant version for the set of classical equations (1). However, it implies in generalizing the time derivative of the action appearing in the first of the equations (1). A simple way to implement it could be by extending this time derivative of $S$ to a divergence of a certain auxiliary four-vector $s^{\mu}$, which was introduced by hand. For $s^{\mu}$ the classical action would be associated somehow with its component $s^{0}$, as the authors briefly discuss in the paragraph just below the equations (8). On the other hand $\lambda^{\mu}$ is just a backgroung four-vector, playing the role of a coupling parameter associated with the dependence of the gravitational lagrangian upon the action. As a first attempt, it is assumed constant, although it could be coordinate-dependent in a more general scenario. One expects that the symmetry of the problem which is being studied helps us to define a specific form for this four-vetor. Besides, different from $s^{\mu}$, which disappears during the variation of the action, $\lambda^{\mu}$ survives and appears in the field equations (see below) as the actual parameter of this model, encoding the dissipative properties that this gravitational theory will manifest. Therefore, any observable quantity of this theory shall be given in terms of $\lambda^{\mu}$.

The resulting field equations are given by \cite{lazo},
\begin{eqnarray}
\label{ei}
R_{\mu\nu} + K_{\mu\nu} - \frac{1}{2}g_{\mu\nu}(R + K) = 8\pi G T_{\mu\nu},
\end{eqnarray}
where $R_{\mu\nu}$ and $T_{\mu\nu}$ are the Ricci and the Hilbert stress-energy tensors, respectively, and 
\begin{eqnarray}
K_{\mu\nu} = \lambda_\rho \Gamma^\rho_{\mu\nu} - \frac{1}{2}\biggr(\lambda_\mu \Gamma^\rho_{\nu\rho} + \lambda_\nu \Gamma^\rho_{\mu\rho}\biggl)
\end{eqnarray}
is a tensor related to viscous geometric dissipations. 

Since the Bianchi identities are still valid, when the divergence of (\ref{ei}) is taken, some relations involving the tensor $K_{\mu\nu}$, its trace and matter sector are obtained. One possibility is to suppose that the gravitational coupling is not constant \cite{lazo}. However, this implies to consider another field responsible for the evolution of $G$, as in the Brans-Dicke theory \cite{BD}. Another possibility is to consider that the divergence of the energy-momentum tensor is not zero, somehow as in the Rastall theory \cite{Rastall}, or as in the Brans-Dicke theory reformulated in the Einstein frame through a conformal transformation.
In what follows we will consider that $G$ is constant. In this case, the usual conservation equations are replaced by,
\begin{eqnarray}
{K^{\mu\nu}}_{;\mu} - \frac{K^{;\nu}}{2} = 8\pi G {T^{\mu\nu}}_{;\mu}.
\end{eqnarray}
This relation complement the field equations (\ref{ei}). In contrast to the Rastall theory, however, now the non-conservation of the energy-momentum tensor has a geometrical origin, with a basis on a variational principle.

Finally, it is important to note that the introduction of a preferential direction given by the coupling vector $\lambda_\mu$ breaks the space-time symmetry resulting in the existence of preferred reference frames, while it still remains compatible with the main principles of the general theory of relativity. Moreover, this feature should be expected in a dissipative theory since the non-conservation of energy-momentum tensor is directly related to the space-time symmetry broken. The relation between energy-momentum non-conservation and space-time symmetry broken by the introduction of a preferred referential frame is also found in the gravitational aether scenario introduced in the context of the cosmological constant problem \cite{aether,Wondrak}. However, while in the gravitational aether theory the space-time symmetry is broken by the introduction of a vector field, the "aether", on the matter side of Einstein field equations, in the present work the symmetry is broken by the geometric coupling $\lambda_\mu$.

\section{Background equations}

Choose the flat Friedmann-Lema\^{\i}tre-Robertson-Walker metric:
\begin{eqnarray}
ds^2 = dt^2 - a(t)^2[dx^2 + dy^2 + dz^2].
\end{eqnarray}
Moreover, let us choose an {\it Ansatz} for $\lambda_\mu$:
\begin{eqnarray}
\lambda_0 &=& const. \neq 0,\\
\lambda_i &=& 0.
\end{eqnarray}

We find:
\begin{eqnarray}
R_{00} &=& - 3\frac{\ddot a}{a} = -3(\dot H + H^2), \\
R_{ij} &=& (a\ddot a + \dot a^2)\delta_{ij} = (\dot H + 3H^2)a^2 \delta_{ij},\\
R &=& - 6\biggr(\frac{\ddot a}{a} + \frac{\dot a^2}{a^2}\biggl) = - 6(\dot H + 2H^2),\\
K_{00} &=& - 3\lambda_0 \frac{\dot a}{a} = - 3\lambda_0H,\\
K_{ij} &=& \lambda_0a\dot a\delta_{ij} = \lambda_0 Ha^2\delta_{ij},\\
K &=& - 6\lambda_0\frac{\dot a}{a} = - 6 \lambda_0 H.
\end{eqnarray}

The energy-momentum tensor $T^{\mu\nu}$ are that of a perfect fluid,
\begin{eqnarray}
T^{\mu\nu} = (\rho + p)u^\mu u^\nu - pg^{\mu\nu},
\end{eqnarray}
with the non null components,
\begin{eqnarray}
T^{00} &=& \rho,\\
T^{ij} &=& p a^{-2}\delta^{ij}.
\end{eqnarray}

The equations of motion are now,
\begin{eqnarray}
3H^2 &=& 8\pi G\rho,\\
2\dot H + 3H^2 + 2\lambda_0H &=& - 8\pi Gp,
\end{eqnarray}
where we have defined,
\begin{eqnarray}
H = \frac{\dot a}{a}.
\end{eqnarray}

On the other hand, we will consider that the Bianchi identities imply,
\begin{eqnarray}
K^{\mu\nu}_{;\mu} - \frac{K^{;\nu}}{2} = 8\pi GT^{\mu\nu}_{;\mu} &=& 0.
\end{eqnarray}
This implies, using the previous component, the equation,
\begin{eqnarray}
8\pi G\{\dot\rho + 3H(\rho + p)\} = - 6\lambda_0H^2.
\end{eqnarray}

Defining,
\begin{eqnarray}
\frac{2}{3}\lambda_0 = - 8\pi G\xi_0,
\end{eqnarray}
the ensemble of equations take the following form:
\begin{eqnarray}
3H^2 &=& 8\pi G\rho,\\
2\dot H + 3H^2 &=& - 8\pi G(p - 3\xi_0H),\\
\dot\rho + 3H(\rho + p - 3\xi_0H) &=& 0.
\end{eqnarray}

Remember that the bulk viscosity (Eckart's theory) leads to a pressure,
\begin{eqnarray}
p^* &=& p - \xi(\rho)u^\mu_{;\mu} \nonumber\\
&=& p - 3\xi(\rho)H.
\end{eqnarray}
Hence, the previous construction corresponds to a constant bulk viscosity coefficient:
\begin{eqnarray}
\xi(\rho) = \xi_0.
\end{eqnarray}
The bulk viscosity coefficient $\xi_0$ must be positive, implying that $\lambda_0 < 0$ in order to retain the analogy. In fact there is a more fundamental reason to impose a negative sign for $\lambda_0$: the entropy production observed in this scenario, an extra aspect which reinforce the resemblance to the non-causal viscous model. 
Following the procedure properly detailed in \cite{winfried,roy} one finds the time evolution of the specific entropy 
\begin{equation}
\label{entropy} 
\dot{s}=-\frac{2\lambda_{0}\rho}{nT},
\end{equation}
where $T$ and $n$ are the temperarature and the particle number density, respectively. So, $\lambda_0$ is obliged to have necessarily negative sign, in order to ensure a non-negative entropy rate production predicted by the sencond law of thermodynamics. 

\section{Perturbed equations}

Now we shall perturb the model described in the previous section, by introducing small fluctuations around the background metric:
\begin{eqnarray}
\tilde g_{\mu\nu} = g_{\mu\nu} + \delta g_{\mu\nu},
\end{eqnarray}
where $\tilde g_{\mu\nu}$ is the inhomogeneous, metric $g_{\mu\nu}$ is the background metric and $\delta g_{\mu\nu}$ represents the fluctuation around it.
From now on, we will note,
\begin{eqnarray}
h_{\mu\nu} \equiv \delta g_{\mu\nu}.
\end{eqnarray}
Due to the inverse metric relation,
\begin{eqnarray}
g^{\mu\rho}g_{\rho\nu} = \delta^\mu_\nu,
\end{eqnarray}
we have,
\begin{eqnarray}
\delta g^{\mu\nu} = - h^{\mu\nu}, \quad h^{\mu\nu} = g^{\mu\rho}g^{\mu\sigma}h_{\rho\sigma}.
\end{eqnarray}

We will work in the synchronous coordinate condition:
\begin{eqnarray}
h_{\mu0} = 0.
\end{eqnarray}
It comes out more convenient, using the synchronous coordinate condition, to rewrite the field equations as,
\begin{eqnarray}
R_{\mu\nu} + K_{\mu\nu} &=& 8\pi G\biggr[T_{\mu\nu} - \frac{1}{2}g_{\mu\nu}T\biggl],\\
K^{\mu\nu}_{;\mu} - \frac{K^{;\nu}}{2} &=& 8\pi G T^{\mu\nu}_{;\mu}.
\end{eqnarray}

The perturbed field equations are:
\begin{eqnarray}
\delta R_{\mu\nu} + \delta K_{\mu\nu} &=& 8\pi G\biggr[\delta T_{\mu\nu} - \frac{1}{2}h_{\mu\nu}T - \frac{1}{2}g_{\mu\nu}\delta T\biggl],\\
\delta (K^{\mu\nu}_{;\mu}) - \delta\biggr(\frac{K^{;\nu}}{2}\biggr) &=& 8\pi G \delta (T^{\mu\nu}_{;\mu}).
\end{eqnarray}

The pertubation of the Ricci and $K_{\mu\nu}$ tensors read,
\begin{eqnarray}
\delta R_{\mu\nu} &=& \partial \delta\Gamma^\rho_{\mu\nu} -\partial_\nu\delta\Gamma^\rho_{\mu\rho} + \Gamma^\rho_{\rho\sigma}\delta\Gamma^\sigma_{\mu\nu}\nonumber\\
&-& \Gamma^\sigma_{\rho\mu}\delta\Gamma^\rho_{\sigma\nu} - \Gamma^\sigma_{\nu\rho}\delta\Gamma^\rho_{\mu\sigma} + \Gamma^\sigma_{\mu\nu}\delta\Gamma^\rho_{\sigma\rho},\\
\delta K_{\mu\nu} &=& \lambda_\alpha \delta\Gamma^\alpha_{\mu\nu} - \frac{1}{2}\biggr(\lambda_\mu\delta\Gamma^\alpha_{\nu\alpha} + \lambda \delta\Gamma^\alpha_{\mu\alpha}\biggl),
\end{eqnarray}
where we have supposed that $\lambda_\alpha$ is constant. In the above expressions, the perturbation of the Christoffel symbol reads,
\begin{eqnarray}
\delta\Gamma^\rho_{\mu\nu} = \frac{1}{2}g^{\rho\sigma}\biggr(\partial_\mu h_{\nu\sigma} + \partial_\nu h_{\mu\sigma} - \partial_\sigma h_{\mu\nu} - 2\Gamma^\lambda_{\mu\nu}h_{\sigma\lambda}\biggl).
\end{eqnarray}

The relevant components of the perturbed Ricci tensor for the study of the scalar modes are:
\begin{eqnarray}
\delta R_{00} &=& \frac{\ddot h}{2} + H\dot h, \\
\delta R_{0i} &=& \frac{1}{2}\biggr(\partial_i \dot h - \partial_k\bar h_{ki}\biggl),
\end{eqnarray}
where,
\begin{eqnarray}
h \equiv \frac{h_{kk}}{a^2}, \quad \bar h_{ij} \equiv \frac{h_{ij}}{a^2}.
\end{eqnarray}

The non-null components of the perturbed $K_{\mu\nu}$ tensor are:
\begin{eqnarray}
\delta K_{00} &=& \lambda_0\frac{\dot h}{2}, \\ 
\delta K_{0i} &=& \frac{\lambda_0}{4}\partial_i h,\\
\delta K_{ij} &=& - \lambda_0 \frac{\dot h_{ij}}{2},
\end{eqnarray}
implying,
\begin{eqnarray}
\delta K &=& \lambda_0\dot h.
\end{eqnarray}
It is useful also to write the perturbations of $K^{\mu\nu}$ (the contravariant form):
\begin{eqnarray}
\delta K^{00} &=& \lambda_0\frac{\dot h}{2}, \\ 
\delta K^{0i} &=& - \frac{\lambda_0}{4a^2}\partial_i h,\\
\delta K^{ij} &=& 2\frac{\lambda_0 H}{a^4}h_{ij}- \frac{\lambda_0}{a^4} \frac{\dot h_{ij}}{2}.
\end{eqnarray}

The non-null components of the energy-momentum tensor are:
\begin{eqnarray}
\delta T^{00} &=& \delta\rho, \\
\delta T^{i0} &=& (\rho + p)\delta u^i, \\
\delta T^{ij} &=& \frac{1}{a^4}\biggr(ph_{ij} + \delta p\,a^2\delta_{ij}\biggl),
\end{eqnarray}
implying,
\begin{eqnarray}
\delta T = \delta\rho - 3\delta p.
\end{eqnarray}

In computing all these expressions we have used, of course, the synchronous coordinate condition and the fact that $\lambda_0$ is constant.

The final set of perturbed equations are:
\begin{eqnarray}
&&\ddot h + (2H + \lambda_0)\dot h = 3H^2(1 + 3v_s^2)\delta, \\
&&\partial_i\dot h + \frac{\lambda_0}{2}\partial_i h - \partial_k\dot{\bar h}_{ki} = - 6H^2(1 + \omega)\delta u^i a^2,\\
&&2H\lambda_0\dot h - \frac{\lambda_0}{4a^2}\nabla^2h = 3H^2\biggr\{\dot\delta + [3H(v_s^2 - \omega) - 2\lambda_0]\delta
+ (1 + \omega)\biggr(\theta - \frac{\dot h}{2}\biggl)\biggl\},\\
\frac{\lambda_0}{4a^2}\partial_i\dot h &-& \frac{3}{4} H\frac{\lambda_0}{a^2}\partial_i h - \frac{\lambda_0}{2a^2}\partial_k\dot{\bar h}_{ki}= 
3H^2\biggr\{(1 + \omega)\delta\dot u^i + [(1 + \omega)\frac{\dot\rho}{\rho}+5(1 + \omega)H]\delta u^i + \frac{v_s^2}{a^2}\partial_i\delta\biggl\}.
\end{eqnarray}

In these expressions, we have defined,
\begin{eqnarray}
\delta &=& \frac{\delta\rho}{\rho},\\
\theta &=& \partial_k\delta u^k,\\
\omega &=& \frac{p}{\rho}, \\
v_s^2 &=& \frac{\delta p}{\delta\rho}.
\end{eqnarray}

Using the background relations,
\begin{eqnarray}
&&3H^2 = 8\pi G\rho,\\
\dot\rho &=& - 2\lambda_0 - 3H(1 + \omega),
\end{eqnarray}
defining 
\begin{eqnarray}
\biggr(\frac{\partial_k\partial_l h_{kl}}{a^2}\biggl)^. = g,
\end{eqnarray}
and performing a Fourier mode decomposition, we have the following set of equations:
\begin{eqnarray}
&&\ddot h + (2H + \lambda_0)\dot h = 3H^2(1 + 3v_s^2)\delta, \\
&&k^2(\dot h + \frac{\lambda_0}{2}h) + g = 6H^2(1 + \omega)\theta a^2,\\
&&2H\lambda_0\dot h + \frac{\lambda_0}{4a^2}k^2h = 3H^2\biggr\{\dot\delta + [3H(v_s^2 - \omega) - 2\lambda_0]\delta + (1 + \omega)\biggr(\theta - \frac{\dot h}{2}\biggl)\biggl\},\\
&-& \frac{\lambda_0}{4a^2}\biggr\{k^2[\dot h - 3 Hh] + 2g\biggl\} = 
3H^2\biggr\{(1 + \omega)\dot\theta+(1 + \omega)[- 2\lambda_0 + (2 - 3\omega)H]\theta - k^2\frac{v_s^2}{a^2}\delta\biggl\}.
\end{eqnarray}

On the other hand the background equations admit (for the one fluid case) the analytical solution,
\begin{eqnarray}
a = \biggr\{ - c\biggr[e^{- \lambda_0 (t - t_0)} - 1\biggl] + 1\biggl\}^\frac{2}{3(1 + \omega)},
\end{eqnarray}
where $c$ is a constant and $t_0$ is the present time.

For the zero pressure case ($\omega = v_s^2 = 0$) the system of perturbed equations reduces to,

\begin{eqnarray}
&&\ddot h + (2H + \lambda_0)\dot h = 3H^2\delta, \\
&&\dot\delta - 2\lambda_0\delta + \biggr(\theta - \frac{\dot h}{2}\biggl)=\frac{2\lambda_0}{3H}\dot h + \frac{\lambda_0}{12a^2}\frac{k^2}{H^2}h,\\
&&\dot\theta + [ 2H- \lambda_0]\theta =\frac{\lambda_0 k^2}{12 H^2 a^2}[\dot h + (3H + \lambda_0)h].
\end{eqnarray}

We can promote a direct comparison between the above set of equations with the case of a single viscous fluid

\begin{eqnarray}
&&\ddot h + \biggr(2H - \frac{H_0\bar\xi_0}{2}\biggl)\dot h = 3H^2\delta - H_0\bar\xi_0\theta,\\
&&\dot\delta + H_0 \bar\xi_0\delta + \biggr(1-\frac{2 H_0 \bar\xi_0}{3 H}\biggl)\biggr(\theta - \frac{\dot h}{2}\biggl) = 0, \\
&&\dot\theta + \biggr(2 H + \frac{H_0\bar\xi_0}{2}\biggl)\theta = \frac{H_0\bar\xi_0 k^2}{6\dot H a^2}\biggr(\theta - \frac{\dot h}{2}\biggl).
\end{eqnarray}

In the above equations $(76)-(78)$ the bulk viscous parameter has been redefined as the dimensionless parameter $\bar\xi_0 = 24 \pi G \xi_0 / H_0$. It is worth noting that making $\lambda_0 =0$ the set $(73)-(75)$ coincides with the equations $(76)-(78)$ for $\xi_0=0$. This corresponds to the pressureless Cold Dark Matter (CDM) case.

We solve numerically the above systems of equations in order to obtain the evolution of the linear density contrast $\delta$. In Fig.1 we show the behavior for the CDM model ($\delta\sim a$) in the red line. For the geometrical model we fix the parameter $\lambda_0 =-0.001$ (in $H_0$ units) in the black dashed line of the left panel. In the right panel we adopt $\lambda_0=-0.01$. The equivalent (at background level) viscous models have the bulk viscous parameter $\bar\xi_0=+0.002$ (left panel) and $\bar\xi_0=+0.02$ (right panel). The growth behavior for such bulk viscous matter is seen in the blue line in both panels. Both curves have the same initial conditions which is equivalent to a $k=0.2 h Mpc^{-1}$ mode deep in the matter dominated epoch. As expected such scale just recently entered the nonlinear regime $\delta\sim 1$.

The curves for the viscous model show the expected pathological behavior as already shown in Refs. \cite{Velten:2011bg} (see also \cite{Li:2009mf}
). Structure growth is highly suppressed in pure viscous cosmologies. On the other hand, the geometrical model follows the CDM behavior. Indeed by increasing the magnitude of $\lambda_0$ a suppression is expected. Let us recall that a slight suppression of the matter clustering on small scales can be important to alleviate the tensions of the $\Lambda$CDM model due to an excess of power verified in numerical simulations \cite{sats}.

In the same spirit of several proposals presented in the last decades, this model is an attempt to abandon cosmological constant, by replacing it by an another gravitational theory. In the present case the cosmic acceleration process would be in charge of the dissipative effects naturally emerging in this new gravity. Dissipative processes are commonly found in the nature, and there is no reason to discard them at all as a possible ingredient in the dynamics of the universe. However, it is reasonable to include them in the cosmology through a viable cosmological model, consistent with the observational data. The triumph of $\Lambda$CDM in this aspect place it as a fiducial model to be compared with. This duty forces us to look at some of the main existing datasets, both at background and at perturbative level, in order to pursue a suitable comparison with the standard scenario and also check the consistency of our model. Such data analysis is going to be performed in a future investigation.

\begin{figure}[t]
\includegraphics[width=0.4\textwidth]{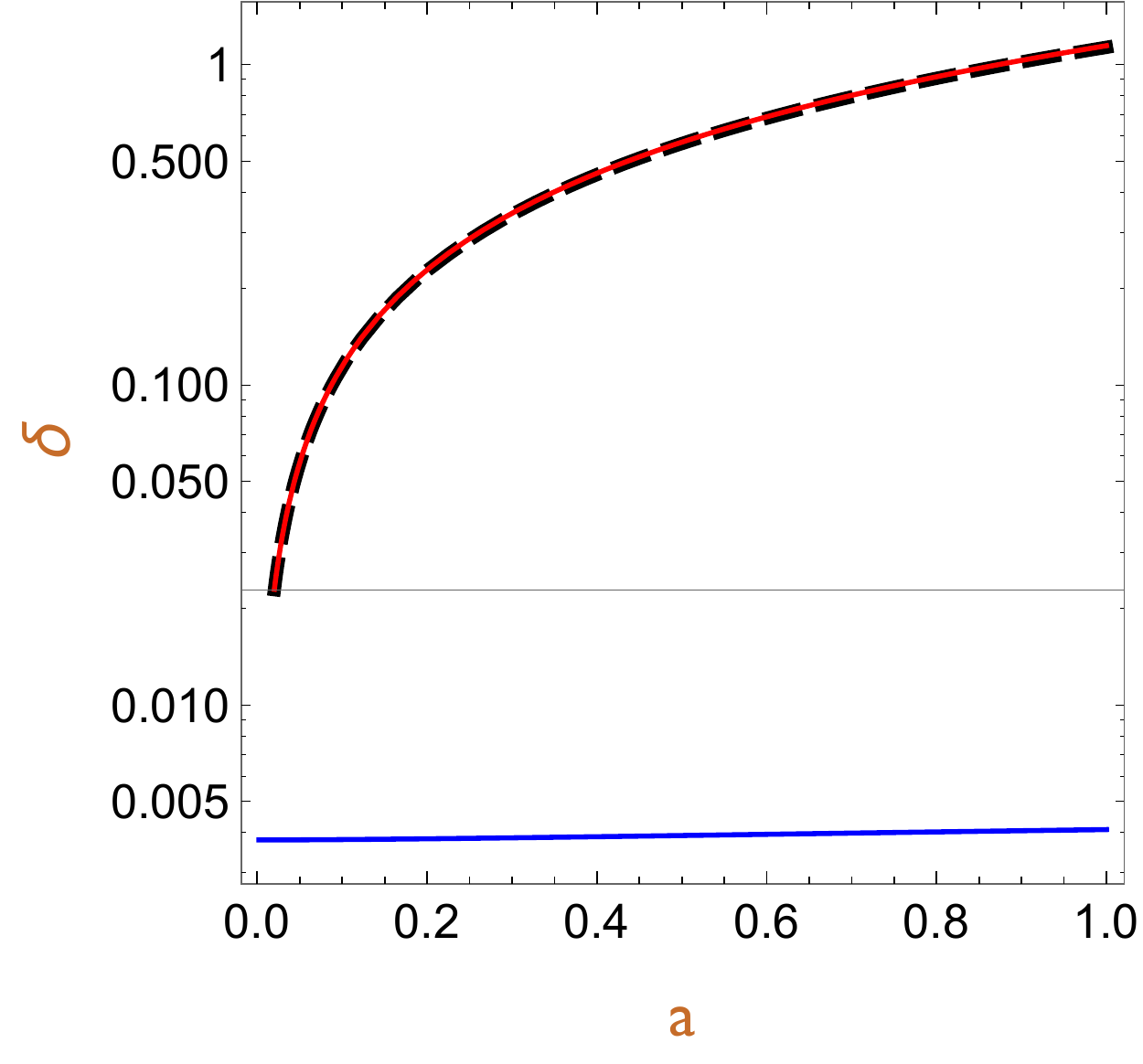}
\includegraphics[width=0.4\textwidth]{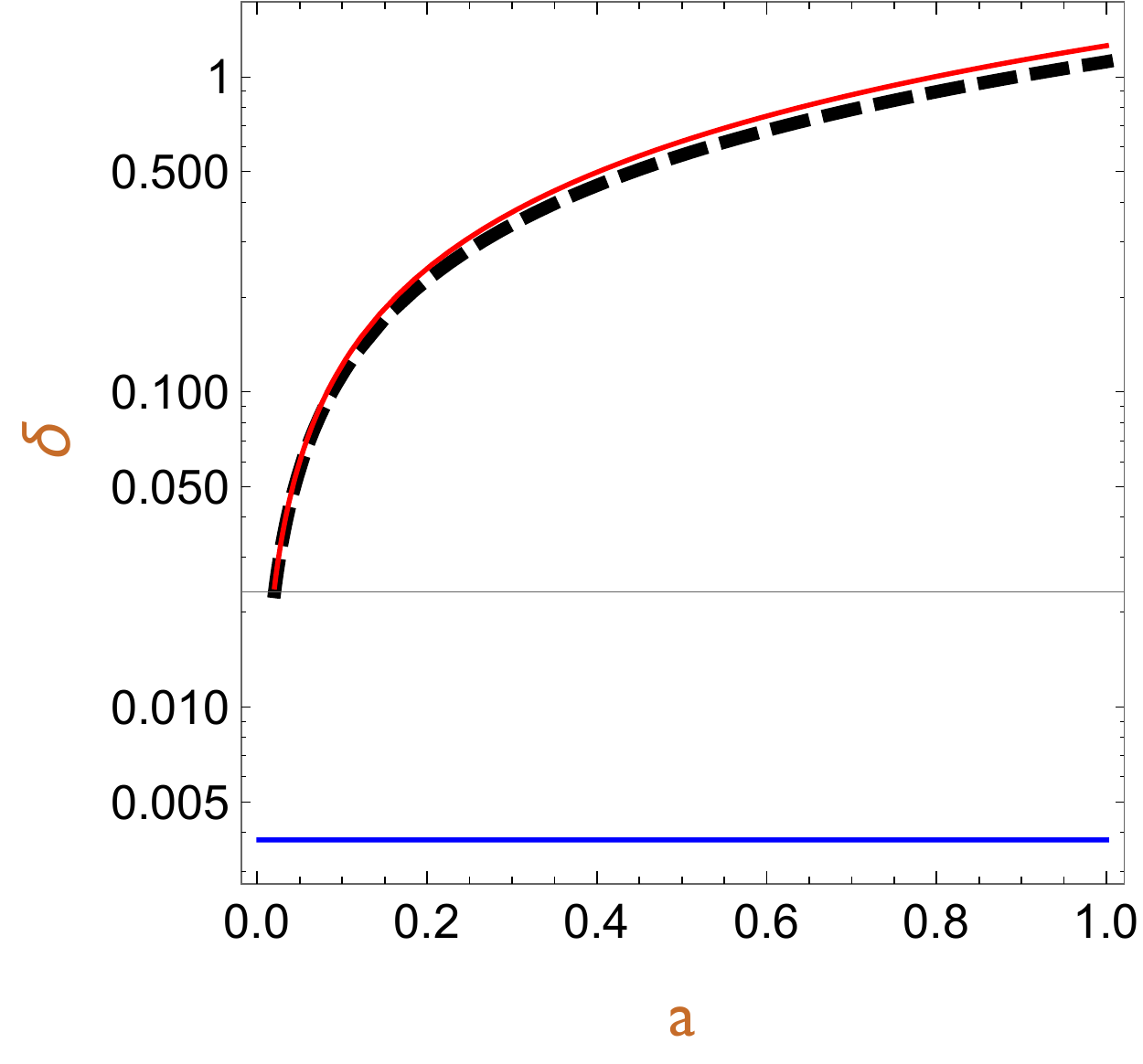}
\caption{Evolution of the matter density contrast as a function of the scale factor $a$. Red line represents the CDM model. For the non-conservative model black-dashed lines are used with values $\lambda_0=-0.001$ (left panel) and $\lambda_0=-0.01$ (right panel). The blue line represents the case in which dark matter possesses bulk viscosity.}
\end{figure}

\section{Final Remarks}

Ref. \cite{lazo} developed the gravitational field equations for a class of theories based on action-dependent Lagrangians. We have studied in this work the flat FLRW cosmology of the resulting theory. Interestingly, we found a deep connection between such formalism and the unified bulk viscous cosmologies (in the Eckart formalism) which have been widely studied in the literature \cite{Fabris:2005ts, Colistete:2007xi, 
	HipolitoRicaldi:2009je, HipolitoRicaldi:2010mf, Fabris:2011wk, Velten:2011bg}. This should be seem as the main result of this contribution. 
	
We have also presented the perturbative dynamics of this model focusing on the scalar matter density fluctuations in the synchronous gauge. Our preliminary analysis shows that the geometrical viscous model does not present the same pathological behavior as the fluid viscous one concerning the growth of cosmic structures. This allows us to promote a proper comparison with matter clustering data in a future work. Such analysis will set the viability of cosmological scenarios based on the new class of action-dependent gravitational theories.

{\bf Acknowledgements:} The authors are grateful to CNPq (Brazil) and FAPES (Brazil) for financial support. We also thank the anonymous referees for their helpful comments and suggestions.


\begin{thebibliography}{}
\bibitem{lazo} M.J. Lazo, J. Paiva, J.T. S. Amaral, and G.S. F. Frederico, 
Phys. Rev. D {\bf 95}, 101501(2017)

\bibitem{her1} G. Herglotz, {\it Ber\"uhrungstransformationen}, Lectures at the
University of G¨ottingen, G\"ottingen, (1930).

\bibitem{her2} R. B. Guenther, C. M. Guenther and J. A. Gottsch, {\it The
Herglotz Lectures on Contact Transformations and Hamiltonian
Systems}, Lecture Notes in Nonlinear Analysis, Vol.
1, Juliusz Schauder Center for Nonlinear Studies, Nicholas
Copernicus University, Tor\"un, (1996)

\bibitem{BD} C. Brans and R.H. Dicke, Phys. Rev. {\bf 124}, 925(1961)

\bibitem{Rastall} P. Rastall, Phys. Rev. D {\bf 6}, 3357(1972); Can. J. Phys. {\bf 54}, 66 (1976)

\bibitem{aether} T. Jacobson and D. Mattingly, Phys. Rev. D {\bf 64}, 024028 (2001).

\bibitem{Wondrak} M. F. Wondrak, {\it The Cosmological Constant and Its Problems: A Review of Gravitational Aether}. In: Hossenfelder S. (eds) Experimental Search for Quantum Gravity. FIAS Interdisciplinary Science Series. Springer, Cham (2018).
[arXiv:1705.06294 [gr-qc]].

\bibitem{winfried} W. Zimdahl, Phys. Rev.  D {\bf 53}, 5483 (1996)
%
\bibitem{roy} R. Maartens and V. M\'endez, Phys. Rev.  D {\bf 55}, 1937 (1997)
\bibitem{Fabris:2005ts} 
  J.~C.~Fabris, S.~V.~B.~Gon\c calves and R.~de S\'a Ribeiro,
  Gen.\ Rel.\ Grav.\  {\bf 38}, 495 (2006)
  doi:10.1007/s10714-006-0236-y
  [astro-ph/0503362]

\bibitem{Colistete:2007xi} 
  R.~Colistete, J.~C.~Fabris, J.~Tossa and W.~Zimdahl,
  Phys.\ Rev.\ D {\bf 76}, 103516 (2007)
  [arXiv:0706.4086 [astro-ph]]
	
\bibitem{HipolitoRicaldi:2009je} 
  W.~S.~Hipolito-Ricaldi, H.~E.~S.~Velten and W.~Zimdahl,
J. Cosmol. Astropart. Phys. {\bf 0906}, 016 (2009)
  [arXiv:0902.4710 [astro-ph.CO]]
	

	
\bibitem{HipolitoRicaldi:2010mf} 
  W.~S.~Hipolito-Ricaldi, H.~E.~S.~Velten and W.~Zimdahl,
  Phys.\ Rev.\ D {\bf 82}, 063507 (2010)
  [arXiv:1007.0675 [astro-ph.CO]]
	
	
\bibitem{Fabris:2011wk} 
  J.~C.~Fabris, P.~L.~C.~de Oliveira and H.~E.~S.~Velten,
  Eur.\ Phys.\ J.\ C {\bf 71}, 1773 (2011)
  [arXiv:1106.0645 [astro-ph.CO]]
	
\bibitem{Velten:2011bg} 
  H.~Velten and D.~J.~Schwarz,
  J. Cosmol. Astropart. Phys. {\bf 1109}, 016 (2011)
  [arXiv:1107.1143 [astro-ph.CO]]
	
	

\bibitem{Li:2009mf} 
  B.~Li and J.~D.~Barrow,
  Phys.\ Rev.\ D {\bf 79}, 103521 (2009)
  [arXiv:0902.3163 [gr-qc]].
 
\bibitem{sats} A. A. Klypin, A. V. Kravtsov, O. Valenzuela, F. Prada, Astrophys. J. {\bf 522}, 82 (1999); B. Moore , S. Ghigna, F. Governato, G. Lake, T. Quinn, J. Stadel, P. Tozzi,  Astrophys. J. {\bf 524}, 19 (1999); V. Springel, J. Wang, M. Vogelsberger, A. Ludlow, A. Jenkins, A. Helmi, J. F. Navarro, C. S. Frenk , S. D. M. White, Mon. Not. Roy. Astron. Soc. {\bf391}, 1685 (2008); J. Diemand, M. Kuhlen, P. Madau, M. Zemp, B. Moore, D. Potter, J. Stadel,  Nature {\bf454}, 735 (2008).
 
    
  
\end{thebibliography}
\end{document}